\documentclass[fleqn,usenatbib]{mnras}

\usepackage{newtxtext,newtxmath}
\usepackage[T1]{fontenc}
\usepackage{ae,aecompl}

\usepackage{graphicx}	
\usepackage{amsmath}	
\usepackage{amssymb}	

\defcitealias{Zuc2015}{Paper~I}
\defcitealias{Zuc2016}{Paper~II}
\defcitealias{Zuc2018}{Paper~III}

\title[PDC Periodogram for Astrometry]{Detection of Periodicity Based on Independence Tests -- IV. Phase Distance Correlation Periodogram for Two-Dimensional Astrometry}

\author[S. Zucker]{
Shay Zucker$^{1}$\thanks{E-mail: shayz@post.tau.ac.il} \\
$^{1}$Department of Geophysics, Raymond and Beverly Sackler Faculty of 
Exact Sciences, Tel Aviv University, Tel Aviv, 6997801, Israel}

\date{Accepted XXX. Received YYY; in original form ZZZ}

\pubyear{2018}

\begin{document}
\label{firstpage}
\pagerange{\pageref{firstpage}--\pageref{lastpage}}
\maketitle

\begin{abstract}
I present an extension of the phase distance correlation periodogram to two-dimensional astrometric data. I show that this technique is more suitable than previously proposed approaches to detect eccentric Keplerian orbits, and that it overcomes the inherent bias of the joint periodogram to circular orbits. This new technique might prove to be essential in the context of future astrometric space missions such as {\it Theia}.
\end{abstract}




\section{Introduction}
\label{sec:intro}

In the three previous papers in this series (\citealt{Zuc2015,Zuc2016,Zuc2018}; hereafter \citetalias{Zuc2015}, \citetalias{Zuc2016} and \citetalias{Zuc2018}), I have introduced a new non-parametric approach to the detection of periodicities in sparse data sets. The first two papers followed the logic of string-length techniques \citep[e.g.,][]{LafKin1965,Cla2002} and presented a way to quantify the dependence between consecutive phase-folded samples (serial dependence), for every trial period. Instead of the classic approach which essentially quantified the linear (Pearson) correlation in those pairs of samples, I proposed to measure the generalized dependence thereof. \citetalias{Zuc2015} and \citetalias{Zuc2016} demonstrated that the new approach was useful for periodicities of the sawtooth type, which included light curves of Cepheids and RR-Lyrae stars, and also radial-velocity curves of eccentric single-lined spectroscopic binaries.

\citetalias{Zuc2018} followed the logic of the Fourier-oriented periodicity-detection methods  such as the popular Lomb--Scargle (LS) technique \citep{Lom1976,Sca1982} and its generalizations \citep[e.g.,][]{Bre2001a,Bre2001b,ZecKur2009}. The starting point was the observation that the LS expression is in fact equivalent to the linear correlation between a real scalar variable and a circular variable\footnote{Here and in the following text, whenever I refer to 'LS' I mean the floating-mean version of LS, introduced by \citet{ZecKur2009}.}, suggested by \citet{Mar1976}. In the context of periodicity search, the circular variable is the phase, calculated according to the trial period. The linear correlation is limited to discover linear relations between variables. I proposed to extend this approach to search for statistical dependence which is not necessarily linear. To this end I proposed to use the {\it distance correlation} -- a dependence measure introduced by \citet{Szeetal2007}. 

Unlike what its name might imply, distance correlation is not a measure of correlation, but of general statistical dependence. Two random variables are independent when their joint probability density function (PDF) is a product of the two marginal PDFs. \citeauthor{Szeetal2007} have introduced the distance correlation as a statistic to quantify the degree of deviation from such independence. As an example, let $X$ be a random variable uniformly distributed between $-1$ and $1$, and let $Y=|X|$. Obviously this is a strong dependence, but the linear correlation in this case will vanish (the $-1$ slope for negative $X$ cancels the $+1$ slope for positive $X$). On the other hand, distance correlation, as a dependence measure, does not vanish in this case (in fact it assumes the value $0.25$).

Distance correlation is gaining popularity in many fields of science \citep{SzeRiz2017}. However, it had not been previously adapted to deal with circular variables. Thus, in \citetalias{Zuc2018}  I have modified distance correlation to apply to circular variables and used this modification to construct a periodicity metric which I dubbed {\it Phase Distance Correlation} (hereby PDC). I have demonstrated that the performance of this new approach in the case of sawtooth periodicities was better than that of the methods I had introduced in the first two papers, while it performed almost equally well as LS in the case of sinusoidal periodicity. 

In the current paper I extend the applicability of PDC to two-dimensional astrometric data.  The most obvious reason for astrometric periodicity of a stellar object is orbital motion caused by the presence of a stellar or planetary companion. Thus, the detection of astrometric periodicity is important in the search of exoplanets.

In the context of the preparations to the launch of the {\it SIM PlanetQuest} astrometric mission (which was eventually canceled), \citet{Catetal2006} proposed an astrometric periodogram, which they dubbed {\it Joint Periodogram}. The joint periodogram consisted of simply adding up the powers of the separate LS periodograms of the two coordinates -- right ascension and declination. In their performance analysis of the joint periodogoram for {\it Sim PlanetQuest}, \citet{Unw2008} admittedly tested this technique under simplified conditions, namely circular face-on planetary orbits. Later, \citet{Bro2009} showed that the performance of the joint periodogram degraded for high eccentricities. The reason is that the LS periodogram is naturally strongly biased towards sinusoidal periodicities. The astrometric signature of circular orbits is two sinusoids in the two coordinates (even if the orbit is inclined). An eccentric orbit no longer projects pure sinusoids on the two dimensions, but involves higher harmonics, which, as a result, degrades the performance of the joint periodogram. Considering multi-harmonic periodogram potentially might alleviate this problem \citep{Sch1996,Bal2009}. Another drawback of the joint periodogram is the dependence on our arbitrary choice of coordinates. Right ascension and declination are convenient for astronomers based on Earth, but they have no fundamental meaning with respect to the target. Thus, the joint periodogram does not represent a universal periodicity metric, but one that depends on an essentially arbitrary choice of coordinates.

\citet{Bosetal2009}, who conducted the ground-based Carnegie Astrometric Planet Search program, proposed as an alternative to use a least-squares periodogram, which would solve, for each trial period, a full 12-parameter astrometric model (including all the Keplerian elements, parallax and proper motion). This approach, in principle, solves the eccentricity problem. However, since it involves fitting for a relatively large number of parameters, it requires more data, as \citeauthor{Bosetal2009} pointed out. This is a serious drawback, since often a decision about follow-up of a target needs to be taken based only on a handful of measurements. Additionally, since it fits simultaneously for many parameters whose nature is very different (angles, times, dimensionless quantities etc.), it is prone to numerical instabilities and biases.

In the next Section I provide a detailed recipe to calculate the astrometric two-dimensional PDC, in Section \ref{sec:perf} I provide some examples and benchmarking, and I conclude in Section \ref{sec:conc} with some further insights.

\section{Astrometric Phase Distance Correlation}
\label{sec:PDC}

As \citet{Szeetal2007} pointed out, an important advantage of distance correlation over Pearson correlation is the fact that the two variables whose dependence is tested need not have the same dimension. Thus the PDC periodogram is readily applicable to two-dimensional data. I hereby repeat the recipe I have described in \citetalias{Zuc2018}, only this time adapted to two-dimensional data:

Let us assume our data set consists of $N$ two-dimensional samples $(\xi_i,\eta_i)$ ($i=1,...,N$), taken at times $t_i$, and we wish to calculate the phase distance correlation for the trial period $P$. In the context of narrow-field, small-angle, differential astrometry, $\xi_i$ and $\eta_i$ would usually mean $\Delta\alpha_i \cos\delta_i$ and $\Delta\delta_i$ -- the right ascension and declination local plane coordinates.

Let us now compute an $N$ by $N$ sample distance matrix, based on the Euclidean distances:
\begin{equation}
a_{ij} = \sqrt{(\xi_i-\xi_j)^2 + (\eta_i-\eta_j)^2}
\end{equation}

In order to compute a phase distance matrix we start by calculating a phase difference matrix:
\begin{equation}
\phi_{ij} = (t_i-t_j) \mod{P}
\end{equation}
which we then convert to a phase distance matrix:
\begin{equation}
\label{eq:phd}
b_{ij} = \phi_{ij}(P-\phi_{ij})\,.
\end{equation}

I have justified this expression for phase distance in the Appendix of \citetalias{Zuc2018}. 

The next step in the calculation is 'double centring' of the matrices $a$ and $b$, i.e.,
\begin{equation}
\label{eq:distances}
A_{ij} = a_{ij}-a_{i\cdot}-a_{\cdot j}+a_{\cdot\cdot} \\
B_{ij} = b_{ij}-b_{i\cdot}-b_{\cdot j}+b_{\cdot\cdot}
\end{equation}
where $a_{i\cdot}$ is the mean of the $i$-th row, $a_{\cdot j}$ is the mean of the $j$-th column, and $a_{\cdot\cdot}$ is the grand mean of the $a$ matrix. Similar definitions apply for the $b$ matrix. $A_{ij}$ and $B_{ij}$ are now the centred distance matrices. We can now finally define the phase distance correlation by:
\begin{equation}
\label{eq:cor}
\mathrm{PDC} = \frac{\sum\limits_{ij}A_{ij}B_{ij}}{\sqrt{(\sum\limits_{ij}A^2_{ij})(\sum\limits_{ij}B^2_{ij})}}
\end{equation}

$\mathrm{PDC}$ is a dimensionless quantity, and it is bounded below by $0$ (complete phase independence), and above by $1$. Higher values mean stronger dependence on phase, and therefore indicate that a periodicity is more likely.

\section{Performance}
\label{sec:perf}

\begin{figure}
\includegraphics[width=\columnwidth,clip=true]{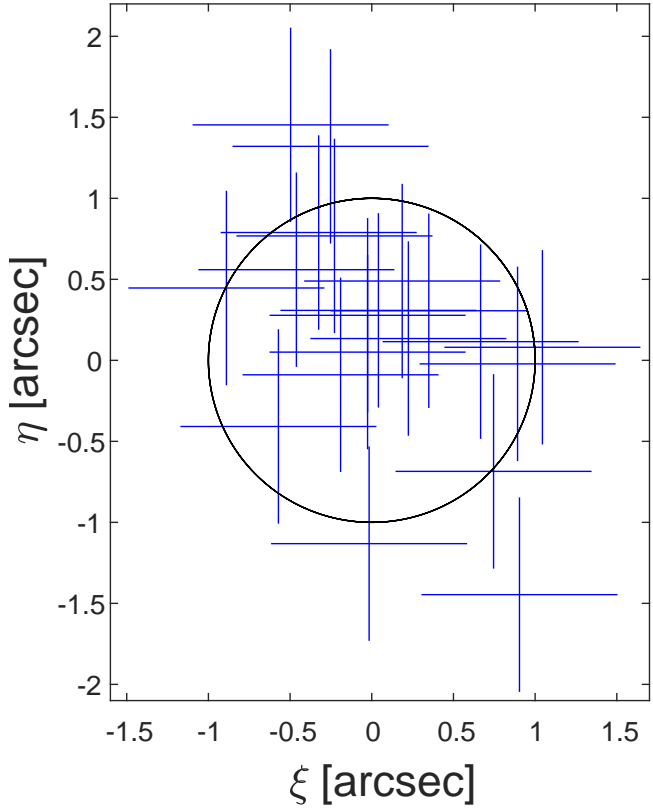}
\caption{An example of two-dimensional astrometric data of a face-on circular Keplerian orbit. The units are arbitrarily assumed to be arcsec. See text for further details.}
\label{fig:faceon2d}
\end{figure}

Figure~\ref{fig:faceon2d} shows simulated astrometric data for a circular face-on Keplerian orbit with an angular semi-major axis of $1\,\mathrm{arcsec}$. The period was assumed to be $30$ days, and I simulated $20$ samples at random times during a period of $100$ days. I assumed an error bar of $0.6\,\mathrm{arcsec}$, and I simulated Gaussian errors in the two dimensions, with no correlation. The Figure presents the astrometric data as blue crosses whose sizes correspond to the error bars, while the orbit itself is shown as solid black line. 

\begin{figure}
\includegraphics[width=\columnwidth,clip=true]{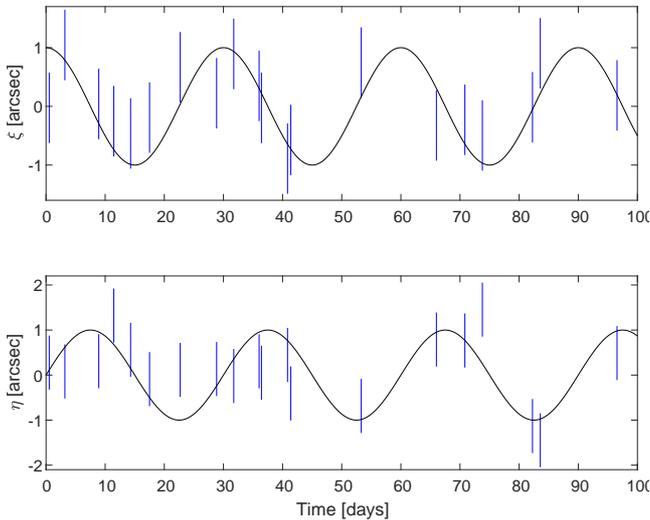}
\caption{The time dependence of the same data shown in Figure~\ref{fig:faceon2d}.}
\label{fig:faceon1d}
\end{figure}

Obviously, in the two-dimensional orbit shown in this Figure the time dependence cannot be demonstrated. Therefore, I show in Figure~\ref{fig:faceon1d} separately the $\xi$ and $\eta$ data as functions of time. The circular orbit clearly translates to pure sinusoids in the two coordinates.

\begin{figure}
\includegraphics[width=\columnwidth,clip=true]{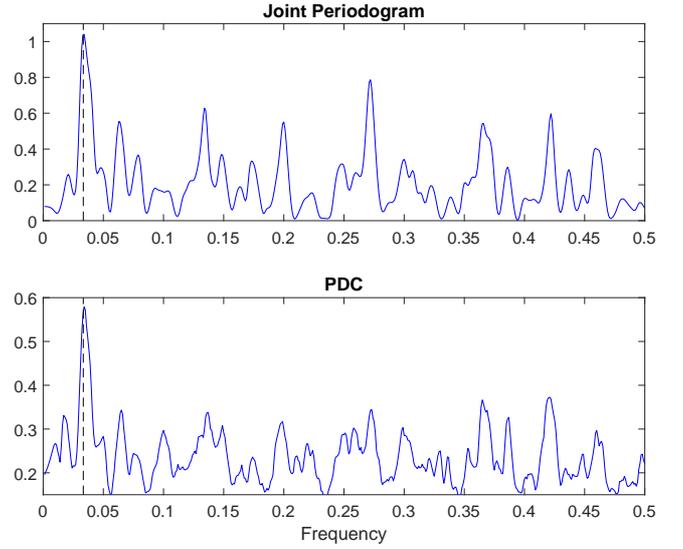}
\caption{The joint periodogram and the PDC periodogram applied to the data shown in Figures \ref{fig:faceon2d} and \ref{fig:faceon1d}. The true frequency is marked by a vertical dashed line.}
\label{fig:faceonper}
\end{figure}

Since the LS periodogram is tailored by definition to detect sinusoidal periodicities, this simple case should not pose any difficulty to the joint periodogram (which is based on LS).  It is still interesting to see how the PDC periodogram performs in this case. Figure~\ref{fig:faceonper} shows the two periodograms applied to this data set. The frequency resolution I used was $0.001\,\mathrm{day}^{-1}$, and the maximum frequency was $0.5\,\mathrm{day}^{-1}$. As expected, the joint periodogram exhibits a clear prominent peak around the known frequency of $(1/30)\,\mathrm{day}^{-1}$, and so does the PDC periodogram.

\begin{figure}
\includegraphics[width=\columnwidth,clip=true]{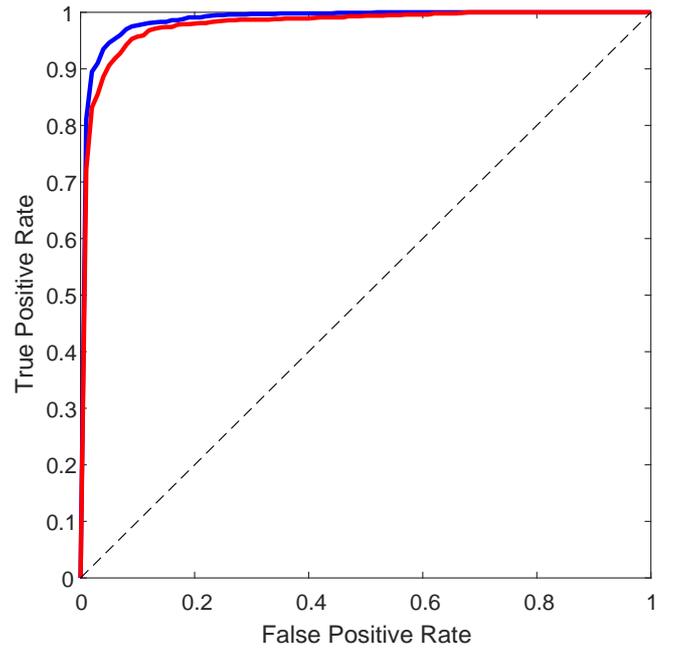}
\caption{ROC curve based on $1000$ simulated cases similar to the one shown in Figures \ref{fig:faceon2d}, \ref{fig:faceon1d} and \ref{fig:faceonper}. The blue line represents the performance of the joint periodogram while the red line represents the performance of the PDC periodogram.}
\label{fig:faceonROC}
\end{figure}

A superficial look at Figure~\ref{fig:faceonper} might create the impression that the PDC periodogram slightly outperforms the joint periodogram, since the peak at the lower panel seems somewhat 'cleaner'. However, Figure~\ref{fig:faceonROC} proves this impression to be wrong. In this Figure I have plotted the resulting  ROC curve, which simply presents the true positive rate against the false positive rate, both determined by the threshold used for deciding on a detection. In the same way I created the ROC curves in \citetalias{Zuc2018}, I simulated $1000$ simulations of pure uncorrelated noise and $1000$ simulations of an orbital signal (with the same characteristics as the example shown above). The detection threshold was applied on the Z-score, like in \citetalias{Zuc2018}. The performance of both periodograms is very good, but there is a small advantage to the joint periodogram, as expected.

\begin{figure}
\includegraphics[width=\columnwidth,clip=true]{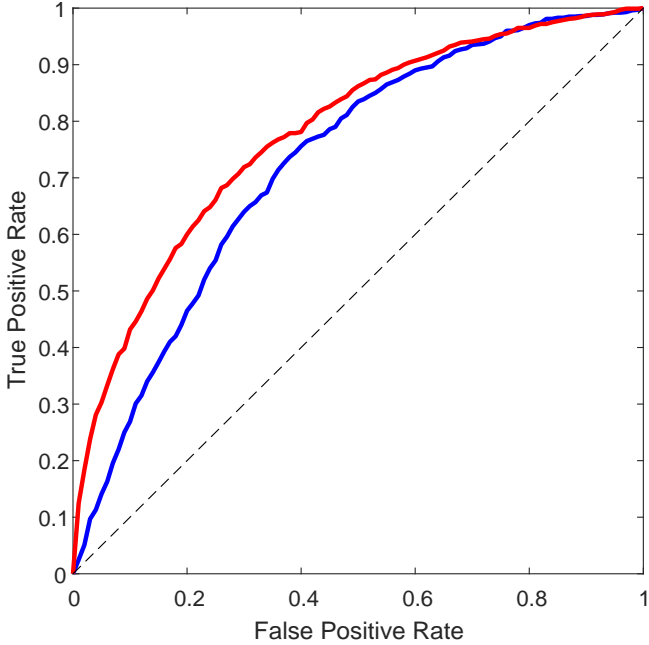}
\caption{ROC curve based on $1000$ simulated cases of an eccentric and inclined orbit (see main text for further details about the simulations).
The blue line represents the performance of the joint periodogram while the red line represents the performance of the PDC periodogram.} 
\label{fig:eccROC}
\end{figure}

Let us now compare the performance of PDC against that of the joint periodogram using simulated data that would challenge the LS-based joint periodogram. To accomplish this I introduced both eccentricity and inclination: I simulated highly eccentric orbits with an eccentricity of $0.99$, and I inclined the orbit relative to the plane of the sky at an inclination of $60\degr$. With eccentricity and inclination, I also had to fix values for the argument of pericentre ($\omega$), and the longitude of the ascending node ($\Omega$). I used the values $\omega=45\degr$ and $\Omega=180\degr$. I also reduced the number of simulated measurements to $10$, and the error bars to $0.1\,\mathrm{arcsec}$. Figure~\ref{fig:eccROC} compares the two periodograms for this configuration.

\begin{figure}
\includegraphics[width=\columnwidth,clip=true]{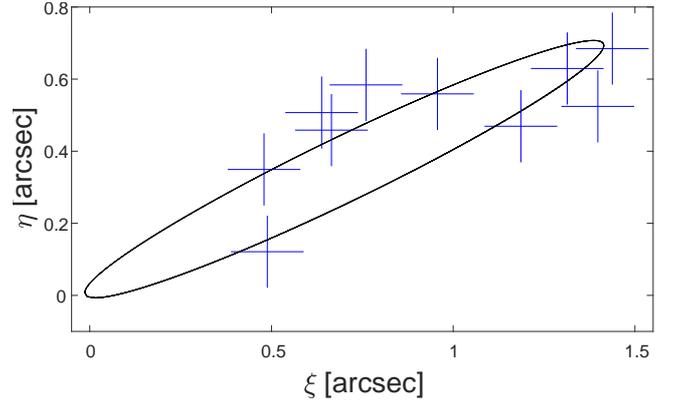}
\caption{An example of two-dimensional astrometric data of an eccentric and inclined Keplerian orbit, one of the positive cases used to create the ROC curve in Figure~\ref{fig:eccROC}.}
\label{fig:ecc2d}
\end{figure}

\begin{figure}
\includegraphics[width=\columnwidth,clip=true]{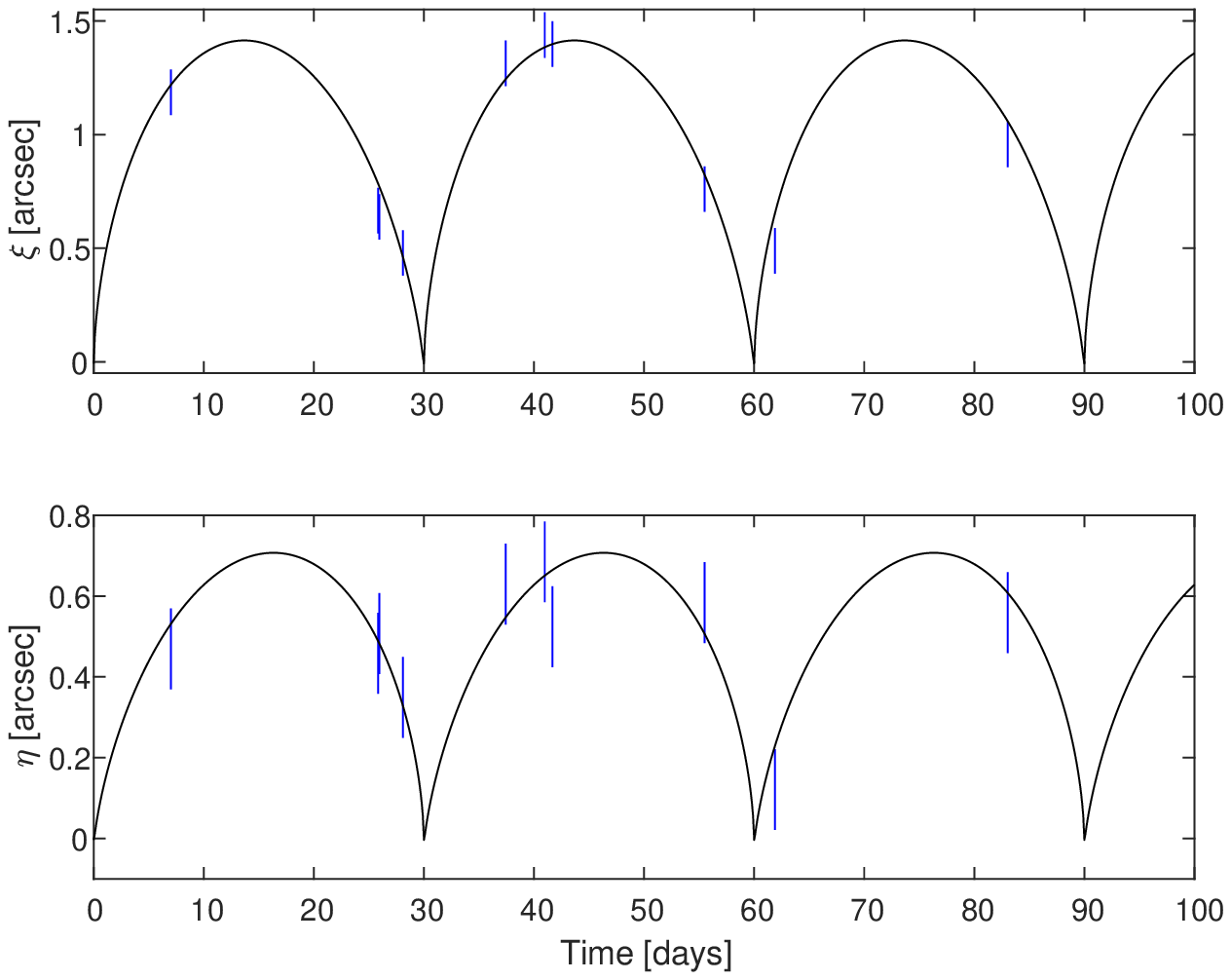}
\caption{The time dependence of the same data shown in Figure~\ref{fig:ecc2d}.}
\label{fig:ecc1d}
\end{figure}

\begin{figure}
\includegraphics[width=\columnwidth,clip=true]{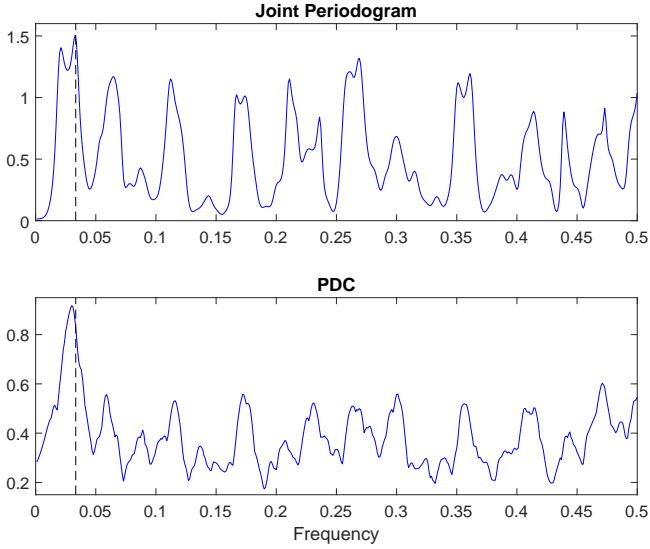}
\caption{The joint periodogram and the PDC periodogram applied to the data shown in Figures \ref{fig:ecc2d} and \ref{fig:ecc1d}. The true frequency is marked by a vertical dashed line.}
\label{fig:eccper}
\end{figure}

Figure~\ref{fig:eccROC} shows that the performance of the two periodograms in this difficult case differs substantially. In fact, at low false-positive rate (the more important part of the curve), say of $0.1$, the true-positive rate increases from $0.30$ for the joint periodogram, to $0.45$ for the PDC -- an increase of $50\%$. In Figures \ref{fig:ecc2d}--\ref{fig:eccper} I show one of the cases which demonstrate the superiority of the PDC over the joint periodogram in this set of simulations. One can clearly see the non-sinusoidal manifestation of the eccentricity, and the way it affects the joint periodogram, rendering it completely ineffective (in this specific case), as well as the clear prominent peak at the correct frequency in the PDC.

\section{Conclusion}
\label{sec:conc}

In the paragraphs above I have shown how the PDC periodogram can be extended to two-dimensional astrometric data, and that it can outperform the joint periodogram in cases of eccentric orbits. The inherent bias of the joint periodogram towards circular orbits (due to the bias towards sinusoids of the LS periodogram in each dimension) is certainly unwanted and needs to be mitigated. PDC may prove to be the solution to this problem. This might be crucial for future planned astrometric space missions, such as {\it Theia} \citep{Maletal2016}. 

It is important at this stage to note that the requirement for two-dimensional data excludes data obtained by scanning astrometry, such as that of {\it Hipparcos}, {\it Gaia} and the like, which are essentially one-dimensional \citep{LinBas2010}, but not {\it Theia}. It is still an open question whether it is possible to modify PDC to data obtained by scanning astrometry, where the scan directions change for every sample.

Besides orbital motion, one can imagine another kind of astrometric periodicity that may arise in cases of 'variability induced movers'. These are caused by the photometric variability of a member of an unresolved pair of targets, which causes an astrometric motion of the centre of light of the pair \citep[e.g.][]{Pouetal2003}. The astrometric signature of this effect will be periodic if the photometric variability is periodic, e.g. if the variable component is a Cepheid. Since photometric variability is not necessarily sinusoidal, this again proves the necessity of an astrometric periodogram that will not be biased to sinusoidal variability in the coordinates.

In summary, this paper shows that the astrometric PDC periodogram should routinely be used, at least to complement the joint periodogram in large (two-dimensional) astrometric surveys, in order to not miss non-sinusoidal astrometric periodicities.

\section*{Acknowledgements}
This research was supported by the ISRAEL SCIENCE FOUNDATION (grant No. 848/16).

\bsp	
\label{lastpage}
\end{document}